# Kinetic Electrostatic Electron Nonlinear (KEEN) Waves and their Interactions Driven by the Ponderomotive Force of Crossing Laser Beams

Bedros Afeyan,[a] K. Won,[a] V. Savchenko,[a] T. W. Johnston,[b] A. Ghizzo,[c] P. Bertrand[c]

[a] Polymath Research Inc., Pleasanton, CA
[b] INRS, Varennes, Quebec, Canada
[c] Universite Henri Poincare, Nancy, France

**ABSTRACT**

*We have found, using 1D periodic Vlasov-Poisson simulations, new nonlinear, nonstationary, stable, long lived, coherent structures in phase space, called kinetic electrostatic electron nonlinear (KEEN) waves. Ponderomotively driven for a short period of time, at a particular frequency and wavenumber, well inside the band gap that was thought to exist between electron plasma and electron acoustic wave frequencies, KEEN waves are seen to self-consistently form, and persist for thousands of plasma periods. KEEN waves are comprised of 4 or more significant phase-locked harmonic modes which persist only when driven sufficiently strongly. They also merge when two or more at different frequencies are driven sequentially. However, the final stable KEEN state that emerges is highly sensitive to their relative order of excitation. KEEN waves also interact quite strongly with electron plasma waves (EPW) especially when their harmonics are close to being resonant with the EPW frequency at the same k. The common assumption that whenever sufficiently large amplitude coherent laser energy is present in an unmagnetized plasma, EPWs and IAWs are the only waves with which the electromagnetic energy can interact coherently may require reconsideration.*

**Keywords:** Kinetic electrostatic electron nonlinear (KEEN) waves, electron plasma waves, Vlasov-Poisson equations, coherent structures in phase space, laser-plasma interactions, ponderomotive force, phase space vortices, chaotic dynamics.

## 1. INTRODUCTION

Since the pioneering work of Bernstein, Green and Kruskal on BGK modes in 1957,[1] it has been quite well known that one could play "plasma designer" and find electrostatic fields that are self-consistent with some vortical phase space structure or some manner of trapped particle velocity distribution function (VDF).[1-3] Of course, the "how do you get there from here?" question had to be gently sidestepped since BGK modes are, strictly speaking, time independent. This has given rise to quite a persistent trend in rarified theoretical studies[4-6] where arbitrary changes to the electron or ion velocity distribution functions are invoked and the troublesome, ubiquitous Landau damping of waves removed with a large nonlinear and arbitrary stroke of the pen. When real laser-plasma interaction experiments[7] revealed that such structures might conceivably be physically relevant, the premium on self-consistently reaching such states and understanding their origins and limitations became even higher.[8]

We decided to take on this challenge by pursuing the most naïve yet sensible limit imaginable. In this limit we simply consider a 1D homogeneous infinite (periodic boundary conditions) Maxwellian electron plasma, with immobile ions. In it a single prescribed wavelength ponderomotive force is imposed (soon after the start of the simulation), with a short temporal envelope (around a 100 inverse plasma frequencies), and a carrier that has a well defined (high) frequency. We followed the evolution of the plasma after this drive was turned off for another 1000 or more inverse plasma frequencies. For fixed (but many) wavenumber(s), we varied the carrier frequency of the drive and to our surprise, found that first, the response was *not* single mode as suggested[5-8] but phase locked *multimode* (higher harmonics were indispensable), and second, that KEEN waves driven by *a wide range of frequencies*, hitherto thought to be inaccessible for coherent collective wave motion of a plasma (in between EPW and EAW frequencies) were sustainable as well. Our studies indicate that the apparent band gap, that is certainly present in linear theory, can be filled with highly nonlinear, strictly kinetic (without a fluid limit, never for a Maxwellian) *not* single mode KEEN waves. The ponderomotive driver has to sufficiently perturb the distribution function locally around its phase velocity in order to give the multimode KEEN wave a chance to survive. This increased local slope reduction of the e⁻ VDF is a prerequisite for trapped particle modes as is well known.[1-8] Oddly enough, however, it is at lower drive amplitudes that the response of the plasma is chaotic and insufficiently organizable. These observations bring into question the hope that a perturbative theory which extends (the rather singular yet formally linear) Van Kampen modes[4] to some nonlinear but manageably small amplitude single mode theory will possibly find use in actual laser-plasma kinetic physics.[6,8] As the drive amplitude is increased, the collective coherent *multimode* response of KEEN waves becomes possible and eventually takes root enough to sustain itself indefinitely (within the context of the V-P 1D description). Furthermore, and most gratifyingly, the sequential introduction of multiple KEEN drives or of KEEN and EPW drives does not destroy these new waves in our simulations. KEEN waves persist in both kinds of "impure" cases, giving more credence to their inherent stability and potential universality. When there are multiple KEEN waves being driven sequentially at different frequencies, they struggle for supremacy and eventually merge to create a unique KEEN wave. But the eventual state that is reached is sensitive to the order in which the KEEN waves were excited. In contrast, when an EPW is present, resonant interactions occur if the frequency ratio between the EPW drive frequency and the KEEN mode's or any of its first few harmonics are sufficiently near one another. This is the case for instance when in natural units (frequency normalized to plasma frequency, wavenumber to the Debye length and thus velocity to the thermal velocity of electrons) $k_{KEEN}$ = 0.26, $\omega_{KEEN}$ = 0.37 and $\omega_{EPW}$ = 1.113. In contrast, $\omega_{KEEN}$ = 0.43 does not produce such a pronounced nonlinear response, as for $\omega_{KEEN}$ = 0.37. As for the range of drive parameters that permit the establishment of single KEEN waves, we have found that they can be formed for all wavenumbers we have tried from 0.1 to 0.55. As a concrete example, for a drive wavenumber of 0.26, and a drive amplitude of 0.2, drive frequencies from 0.3 to 0.6 gave rise to long lasting KEEN waves while the EAW frequency is 0.37. So any frequency within *at least* 50% on either side of the EAW frequency can be used to drive KEENs. Besides, for all these frequencies and even *at* the EAW frequency, what results is a phase locked multimode wave whose second harmonic is twice as large as its fundamental, its third harmonic, just as large as the fundamental and the forth harmonic, only half as large as the fundamental. These phase locked multimode aspects are entirely absent in any EAW prescription of the past.[5-8] In contrast to KEEN waves, the tuning width of EPWs is of order one or two percent or less, typically. This means that the band of frequencies within which KEEN waves may be excited is very wide and certainly *not* the postulated single frequency, single mode, EAW response, or the well known EPW response[2-4], but something else entirely.

## 2. FORMULATION OF THE PROBLEM

We solve the Vlasov-Poisson set of equations in a uniform plasma with periodic boundary conditions assuming an initially Maxwellian velocity distribution function of width one in natural units.[1-4] The method of solution is the split operator semi-Lagrangian cubic spline interpolation scheme which is quite popular since the pioneering work of Chang and Knorr.[10] We introduce an externally imposed ponderomotive (electrostatic) force driver (PFD) which may be thought of as being due to the beating of two counter-propagating electromagnetic waves.[9] The PFD is simply added to the self-consistent electric field in the Vlasov equation. For a single KEEN wave simulation, the PFD is characterized by an amplitude, $a_{Drive}$, wavenumber, $k_{KEEN}$, which must be a multiple of the simulation domain size, a frequency, $\omega_{KEEN}$, and a temporal envelop. We use the difference between two hyperbolic tangent functions to obtain a smoothly varying hat function with a start up ramp of 10 inverse plasma frequency units, a duration of a hundred, and a ramp down of 10 units as well. We have compared ramp down rates of 10, 50 and a 100 so as to detect any peculiarities associated with the fast case but didn't find any. The drive amplitude that enters the Vlasov equation is the spatial derivative of the ponderomotive potential, which itself is simply the product of the oscillatory velocities of electrons in the E field of each of the two crossing lasers[9] divided by the electron thermal velocity of the plasma squared. By $a_{Drive}$ we mean this (normalized) velocity squared quantity. We find that interesting and stable behavior is obtained for drive amplitudes above 0.15. For a clean KEEN wave, 0.2 and above is preferable. For a green pump laser (527 nm) at 2 x 10$^{15}$ W/cm$^2$ and an orange colored probe (600 nm or so) at a tenth that intensity, and a 0.5 keV electron temperature, we find the drive amplitude to be above 0.23, which qualifies as a good KEEN wave generating drive level. This can easily be achieved on Trident[7] at LANL with ten times higher intensities being available as well. The resolution used in these simulations is 2048 in space and 4096 in velocity for a v range of –6 to 6 and 4 modes in x. The time resolution was 0.05 in inverse plasma frequency units. This was seen to closely resemble 0.01 resolution runs but be significantly different than Δt = 0.1 runs.

## 3. VLASOV SIMULATION RESULTS

We show the late time phase space images of a KEEN wave in Fig. 1. On the left, f is plotted, while on the right, the difference between f and the initial Maxwellian normalized to the maximum value of that difference with a v axis shifted by the phase velocity of the driver field. Fig. 2 shows the density response in Fourier space as a function of time for three drive amplitudes, 0.2, 0.1 and 0.05. Note that four modes have large amplitude responses late in time while twenty or so are involved in the creation process of a KEEN wave (early in time, following the turn off of the drive which is at 130). The first case is driven hard enough that it produces a long lasting and undamped KEEN wave. This is not so for the third case shown in Fig. 2. The second case is for marginal drive. The long time stability issue as a function of drive amplitude is also addressed in Fig. 3. Here we plot the values of the first 6 modes of $f-f_0$ in k space, RMS averaged over velocity (in a velocity coordinate which is shifted by the phase velocity of the driver, with n the mode number):

$$\left\langle \hat{f}^{(n)} \right\rangle (t) = \sqrt{\left\{ \int_{-\overline{v}_{MAX}}^{\overline{v}_{MAX}} \left| \hat{f}^{(n)}(\overline{v},t) \right|^2 d\overline{v} \right\}}.$$

We can thus determine which harmonics survive and are essential pieces of a KEEN wave, *both* from the average over v of each f mode, whose magnitudes are the rho's plotted in Fig. 2, *and* by the velocity space average of each f mode's magnitude squared, as shown in Fig. 3.

Fig. 4 shows the changes that occur to the zero order $e^-$ VDF when the KEEN wave is formed. The anticipated reduction of the slope of the distribution function is seen to arise in these self-consistently generated waves only if the drive amplitude is high enough. Fig. 5 shows the reconstruction of $f - f_0$ from 0, 1, 2, 3 and 4 first spatial modes that gets very close to the full KEEN wave in this 0.2 drive amplitude case. The last frame is the very small difference between the "up to the 4$^{th}$ harmonic approximation" reconstruction and the full solution. Fig 6 shows this more clearly still with the absolute values of the space-time Fourier transformed (fast carriers surgically removed) components of f being plotted as a function of $v-v_\phi$. The few significant coupled harmonics nature of a KEEN wave is apparent. These harmonics are phase locked, as one may expect from their coherent multimode dynamical orbits in phase space.[11] The initial $e^-$ VDF need not be Maxwellian for a KEEN wave to form. In fact, when a weak amplitude drive is applied but the initial $e^-$ VDF is extracted from a stronger drive asymptotic state, KEEN waves are found to arise and persist even for that low drive amplitude. The surprise is that this degree of self-organization is apparently possible in a very high dimensional dynamical system (or field theory) well beyond the few coupled oscillator limit where nonlinear stability is more commonly achieved. We are building a parallel version of the Vlasov-Poisson code used herein so as to explore more efficiently these newly found coherent multimode nonlinear kinetic waves in the traditional "band gap" which was thought to exist in plasma physics and laser-plasma interactions.

## ACKNOWLEDGMENTS

This work was supported by the DOE Grant DE-FG03-03NA00059. We would like to thank D. S. Montgomery for his pioneering and inspirational experiments which gave us the impetus to tackle the EAW existence problem with fully nonlinear, self-consistent and untruncated Vlasov-Poisson simulations. We also would like to thank H. Rose, S. Brunner, E. Williams, W. Manheimer, and W. Kruer for valuable discussions and encouragement.

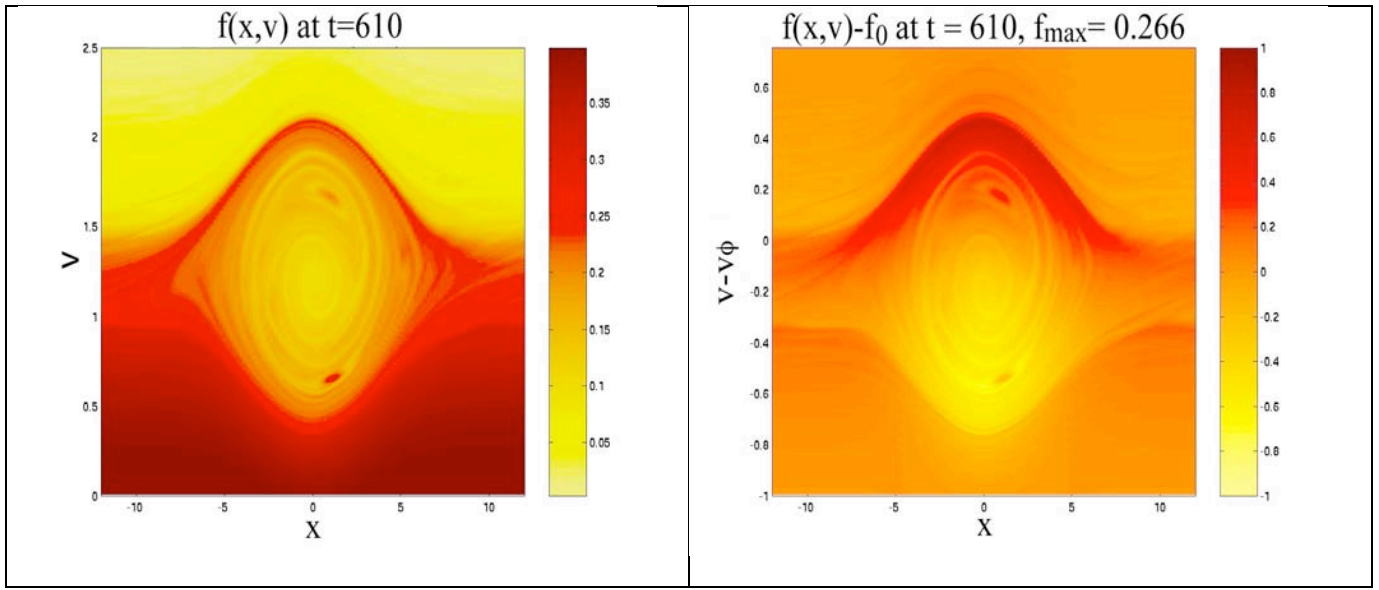

**Figure 1.** Late time phase space snapshots of a KEEN wave ponderomotively driven with an amplitude 0.4, $k_{KEEN}$ = 0.26 and $\omega_{KEEN}$ = 0.37. Fig 1b is also a phase space snapshot where the velocity axis is shifted to be centered with respect to the phase velocity of the driver and where the initial Maxwellian VDF, $f_0$, has been subtracted. Of the four vortices that exist in the computational domain, we show just one for clarity.

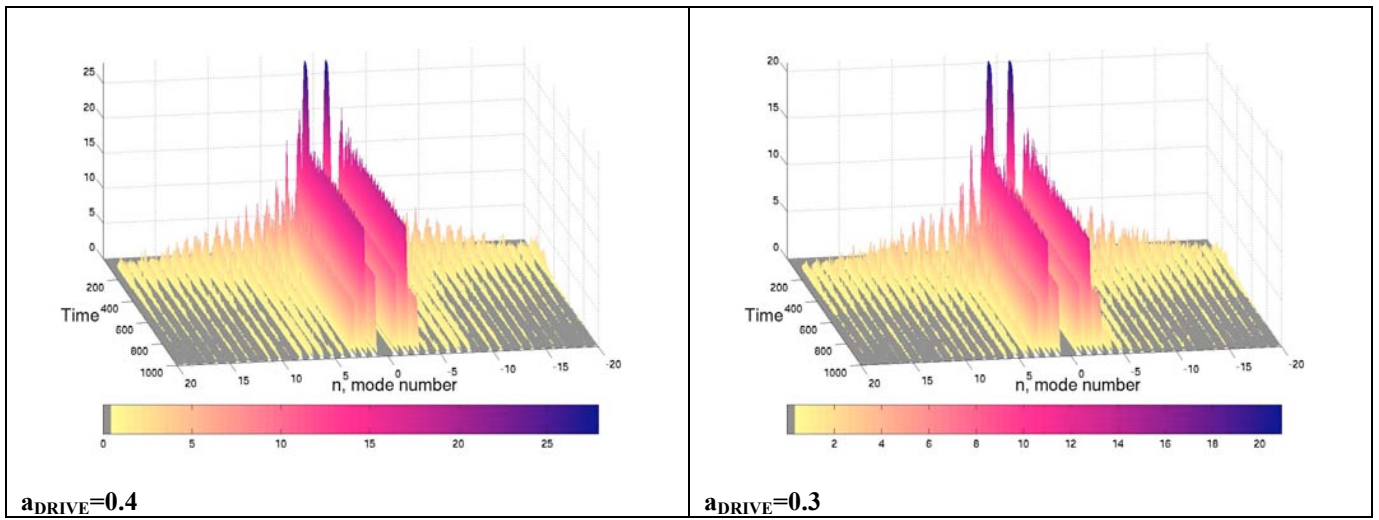

$a_{DRIVE}$=0.4     $a_{DRIVE}$=0.3

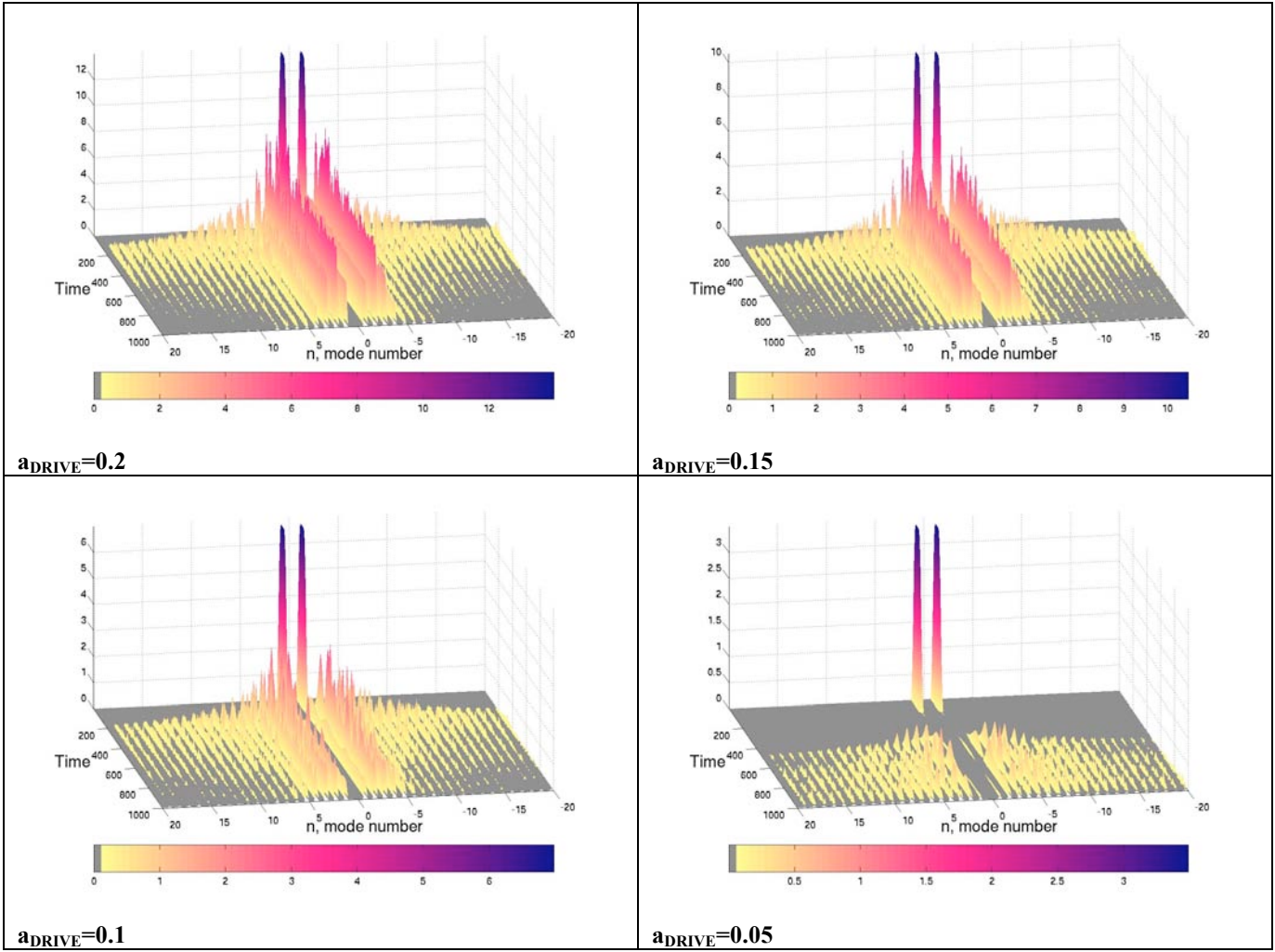

**Figure 2.** The absolute value of the density response as a function of mode number $k/k_{KEEN}$ and time. Six amplitude drive cases are shown side by side. $a_{Drive}$ = 0.4. 0.3, 0.2, 0.15, 0.1 and 0.05. The fourth is a marginal KEEN wave producing case and the last two are abortive ones. Note that around 20 modes are involved with the initial KEEN wave creation process but only 4 or 5 survive with any appreciable amplitude in steady state for the $\omega_{KEEN}$ = 0.37 and $k_{KEEN}$ = 0.26 cases shown.

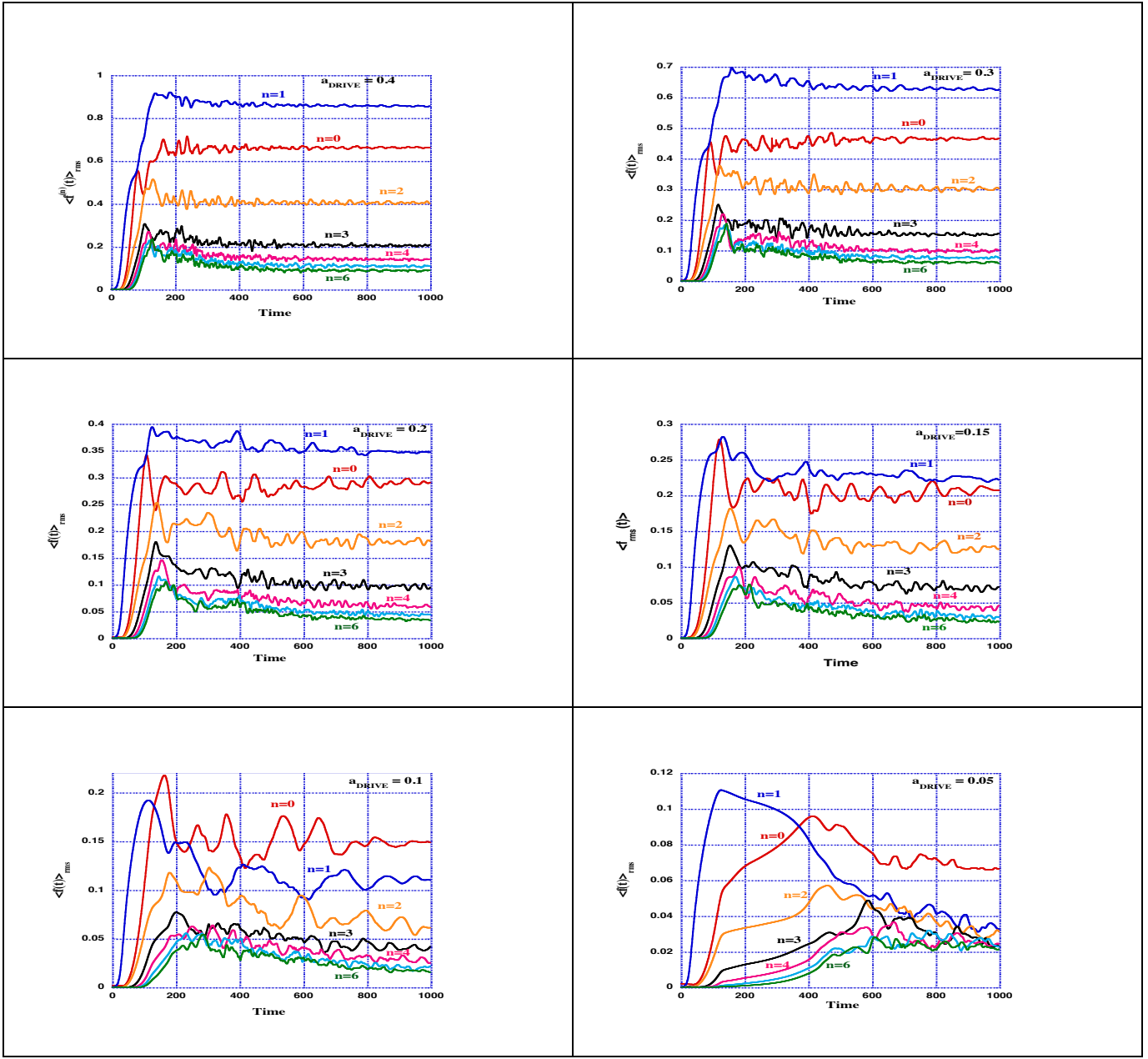

**Figure 3.** The RMS velocity averaged values of the first few (0, 1, …,6) k space modes of KEEN waves driven at amplitudes 0.4, 0.3, 0.2, 0.15, 0.1 and 0.05. Note that at 0.1 and 0.05, the long term trend of the RMS values of f decay while at higher drive amplitudes the time asymptotic values are flat. The higher the drive amplitude, the smoother the RMS response becomes. The multi-mode self-organization is faster and more efficient with higher drive amplitudes.

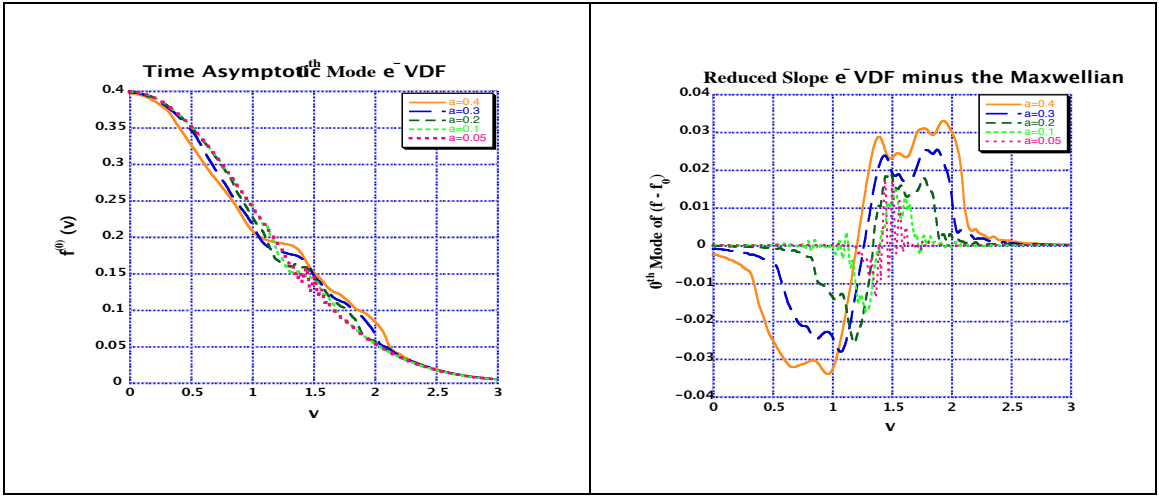

**Figure 4.** The drive amplitude dependence of the time asymptotic form of the non-oscillatory component of the e⁻ VDF of a KEEN wave, spatially averaged f, as well as the spatially averaged f–$f_0$, where $f_0$ is the initial Maxwellian. Drive amplitudes of 0.4, 0.3, 0.2, 0.1 and 0.05 are shown.

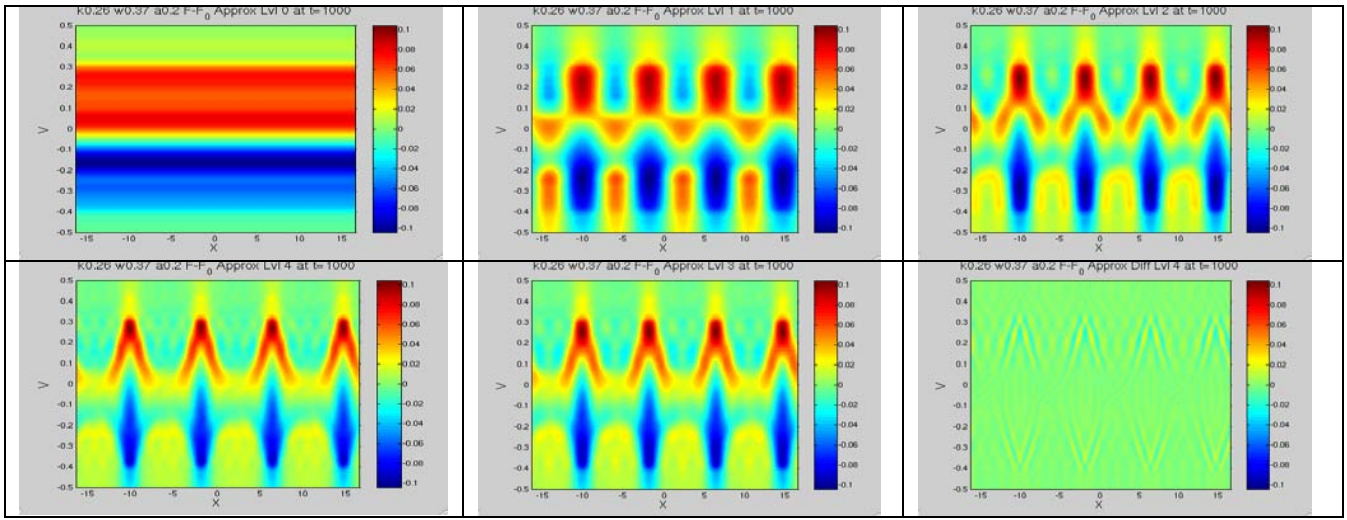

**Figure 5.** Partial mode reconstructions of a large amplitude drive ($a_{Drive}$ = 0.2) KEEN wave. Note that very small corrections are included in modes higher than four.

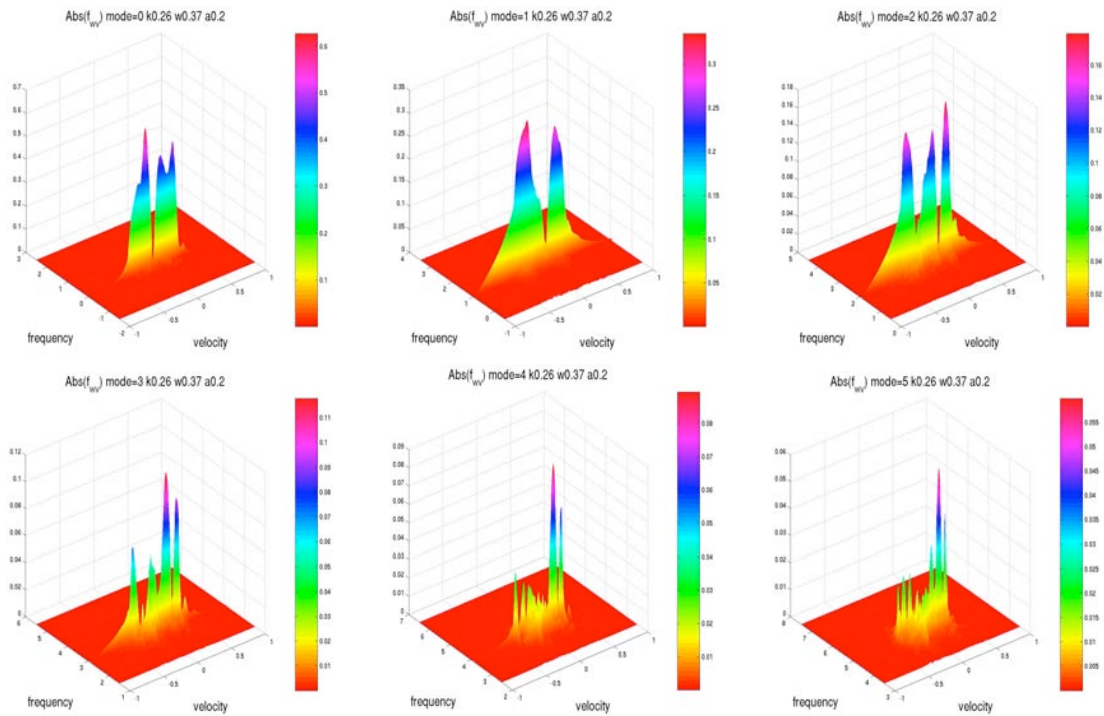

**Fig 6.** The absolute values of the zeroth and the next 5 spatial modes of f vs frequency (normalized to the drive frequency) and velocity (in phase velocity shifted coordinates). This is a drive amplitude 0.2 KEEN wave whose drive frequency was 0.37 and wavenumber, 0.26. Note the very pure harmonics in frequency in the lowest order modes, becoming more and more extended in frequency content as we look at higher and higher modes.